\documentclass[11pt]{article}
\usepackage{graphics}
\usepackage{epsfig}
\usepackage{epstopdf}
\usepackage{amsmath}
\usepackage{array}
\begin{document}

\title{Does Holographic equipartition demand a pure cosmological constant?}

\author{Krishna P B and Titus K Mathew \\ 
Department of Physics, Cochin University of Science and Technology, Kochi-22. \\
 krishnapb@cusat.ac.in; titus@cusat.ac.in}
 \maketitle
\begin{abstract}
 The spacial expansion of the universe could be described as a tendency for satisfying holographic equipartition which inevitably demands the presence of
 dark energy. We explore whether this novel idea proposed by Padmanabhan give any additional insights into the nature of dark energy. In particular, we
 obtain the constraints imposed by the law of emergence on the equation of state parameter, $\omega$. We also present a thermodynamic motivation for the 
obtained constraints on $\omega$. Further, we explicitly prove the feasibility of describing a dynamic dark energy model through the law of emergence.
Interestingly, both holographic equipartition and the entropy maximization demands an asymptotically de Sitter universe with
 $\omega\geq-1$, rather than a pure cosmological constant.
\end{abstract}

 \section{Introduction}

 Various attempts in understanding gravity indicate that, it could be an emergent phenomenon. This paradigm shift was 
 originated  due to the 
 understanding of the connection between gravity and thermodynamics. Following the remarkable discovery of Hawking\cite{Hawking1} and 
 Bekenstein\cite{Bekenstein1} on black hole 
 thermodynamics, Jacobson \cite{Jacobson1} have shown that, the Einstein's equation of gravity is equivalent to the first law of thermodynamics, which connects heat,
 entropy and temperature, $\delta Q=T dS.$. This connection between gravity and thermodynamics was extended to a wide class of gravity theories \cite{Newpaddy}. 
 Inspired from these, Verlinde\cite{Verlinde1} have reformulated gravity as an entropic force, emerged due to the changes in the location of the 
 material bodies. Padmanabhan also have proposed a similar idea at around the same time, in which he derived the Newton's law of gravity using the 
 equipartition law of energy and the relation $S=E/2T,$ where $E$ is the energy of the system\cite{Padmanabhan1}.
 
 Recently, Padmanabhan\cite{Paddy2} carried over this idea of emergence to the extend that the space could also be emergent as time evolves. He postulated that the 
 time evolution of the universe is due to the difference between the degrees of freedom on the horizon and the cosmic components in the 
 bulk region enclosed by the horizon.
The time evolution of the universe will continue until these degrees of freedom becomes equal. This asymptotic equality of the degrees 
of freedom requires 
 a pure de Sitter epoch as the end phase of the universe, which demands the inevitable presence of dark energy. While proposing this 
expansion law, Padmanabhan remarked that, the dark energy component of the universe is not too different from a pure cosmological 
constant. For more investigations on this proposal see \cite{KT1,KT2,MKT,Cai,sheykhi,yang,tu,sumanta3}.

Cosmological constant as dark energy is only a special case for causing the evolution of the universe to a final de Sitter epoch.
Even though cosmological constant provides a nice explanation for the late acceleration of the universe\cite{Komatsu1}, which finally 
tends to a de Sitter epoch, the resulting model, 
the standard $\Lambda$CDM model,
is plagued with the 
so called cosmological constant problem, that the observed and predicted values of the 
cosmological constant,
differ by orders greater than 120\cite{Sahni1,Carroll1,orlando1,Paddy3}. This motivates the introduction of dynamical dark energy 
models, which may be either varying dark energy where both the density and equation of state of dark energy are varying 
or decaying vacuum models 
in which the density corresponding to the so called 'cosmological constant' parameter is varying with cosmic time, but 
equation of state remains constant at -1.
The decaying vacuum models can 
be well motivated using quantum field theory\cite{Shapiro1,Bonanno1,Urban1}. But the occurrence of asymptotic de Sitter 
epoch is common to both these types of models.

More recent observational results are also contradicting the straightforward 
truthfulness of 
the $\Lambda$CDM model. For example, the $H_0$\cite{Riess2,orlando2,orlando3,orlando4} and $\sigma_8$ \cite{Macaulay1,Basilakos2} tensions are 
important enough to reveal the discrepancies of the 
 constant cosmological constant models.  Recent literatures which analyze the cosmological data in a 
critical way\cite{Sola2,Sola3,Zhao1} 
are strongly favoring a slowly varying cosmological parameter 
instead of a pure constant. 
Interestingly such works offer solutions to the previously mentioned tension. As mentioned,
the most interesting thing in these type of decaying vacuum models is that, all of them 
predicts a de Sitter epoch as the end phase of the expanding universe. 
So the idea of a strict cosmological constant could be an 
oversimplification.
These tempting one to check the validity of the 
Padmanabhan's principle, which assumes 
 
In the present work we explore whether the law of emergence demands a pure cosmological constant. More specifically, we analyze whether the law of
emergence give any insights into the nature of dark energy. In addition, we will explicitly prove the validity of the law of emergence in a varying dark
energy model. In the next section, we give a brief review of Padmanabhan's proposal. In section 3, we find the constraints imposed by the holographic
equipartition on the equation of state parameter $\omega$. We present a thermodynamic motivation for these constraints in section 4. In section 5, we 
analyze the feasibility of describing a dynamical dark energy model using the law of emergence. We present our conclusions in section 6.

\section{The law of emergence}

We will begin with a short description of Padmanabhan's emergent paradigm. A pure de Sitter universe obeys holographic equipartition in the form,
\begin{equation}\label{eqn:EP1}
 N_{surf}=N_{bulk}
\end{equation}
Here $N_{surf}$ denotes degrees of freedom on the surface of the Hubble sphere with radius, $r=1/H$ and is given by
\begin{equation}\label{eqn:nsurf1}
 N_{surf}=\frac{4\pi}{l_p^2 H^2}
\end{equation}
where $l_p^2$ is the Planck area that represents one degree of freedom and $H$ is the Hubble parameter. The term $N_{bulk}$ denotes the degrees of freedom 
in the bulk volume bounded by the horizon. This can be obtained by invoking the equipartition law, such that each degree of freedom in the bulk carries an 
energy $\frac{1}{2}k_B T$ where $k_B$ is the Boltzmann constant and $T=H/2\pi$ the Gibbon's Hawking temperature. Consequently $N_{bulk}$ can be defined as,
\begin{equation}\label{eqn:NB}
 N_{bulk}=\frac{\left|E\right|}{\frac{1}{2} k_B T}
\end{equation}
where $\left|E\right|=\left|\rho+3p\right|V,$ the Komar energy within the Hubble volume, $V=\frac{4\pi}{3 H^3}.$ Here $\rho$ and $p$ are the density and pressure of the 
cosmic components. Using equation (\ref{eqn:NB}), equation  (\ref{eqn:EP1}) can be written as, $|E|=\frac{1}{2} k_B T N_{surf}$ which is termed as the holographic equipartition.
 Equation (\ref{eqn:NB}) can also be expressed as,
\begin{equation}\label{eqn:NB2}
 N_{bulk} = -\epsilon \frac{2\left(\rho+3p\right) V}{k_B T}
\end{equation}
where $\epsilon=-1,$ if $\left(\rho+3p\right)>0$, for matter dominated universe and $\epsilon=+1,$ if $\left(\rho+3p\right)<0$, for dark energy dominated phase.
The present observational evidence indicates that, the universe is proceeding towards a pure de Sitter state at which the degrees of 
freedom on the horizon and the bulk are equal. Hence the expansion of the universe is assumed to be driven by the difference in 
the degrees of freedom and  it's dynamical evolution can be expressed as\cite{Paddy2},
\begin{equation}\label{eqn:dvdt1}
 \frac{dV}{dt}=l_p^2\left(N_{surf} - \epsilon N_{bulk} \right).
\end{equation}
This equation generally can be called as the holographic equipartition law.
From this fundamental law, it is possible to derive the Friedmann equations of a flat FLRW universe\cite{Paddy2}. 
For a universe with matter, radiation (rad), and dark energy (de) 
equation (\ref{eqn:dvdt1}) can be expressed as,
\begin{equation}\label{eqn:dvdt2}
 \frac{dV}{dt}= l_p^2 \left(N_{surf} + N_{matter} + N_{rad} - N_{de} \right).
\end{equation}
As the universe is tending towards the de Sitter phase with a constant $H$, the left hand side of the above equation will vanish asymptotically, 
i.e.$\frac{dV}{dt} \to 0;$ when $t\to \infty$. For the right hand side to satisfy the same asymptotic condition, the presence of the dark energy component is essential.

\section{Holographic equipartition and the constraints on $\omega$}
Many authors have checked the
status of Padmanabhan's principle in higher dimensional Einstein's theory of gravity and also in 
alternative theories of gravity, thus ratified it's general validity. 
In all such analysis the dark energy have been incorporated as a pure cosmological constant. Our aim here is to analyze the validity of holographic
equipartition in the presence of dynamical dark energy.

Let us assume that our universe contains a dynamical dark energy component. The surface degrees of freedom
can be expressed as,
\begin{equation}\label{eqn:nsurf2}
 N_{surf}=\frac{4\pi}{l_p^2 H^2}.
\end{equation}
From equation (\ref{eqn:NB2}), we can write the bulk degrees of freedom as,
\begin{equation}\label{eqn:nbulk2}
  N_{bulk}= -\epsilon \frac{(4\pi)^2 \left(1+3\omega \right) \rho}{ 3k_B H^4}.
 \end{equation}
Here $\omega$ is the equation of state parameter defined through the barotropic equation, $p=\omega\rho$. 
 Using the Friedmann equation, 
\begin{equation}
 H^2=\frac{8\pi G \rho}{3},
\end{equation}
and by taking $l_p^2\sim G$, $ k_B =1$, the bulk degrees of freedom can be rewritten as,
\begin{equation}\label{eqn:nbulk3}
  N_{bulk}= -\epsilon \frac{(4\pi) \left(1+3\omega \right) }{ 2 l_p^2 H^2}.
 \end{equation}
According to Padmanabhan's proposal, when the universe satisfies holographic equipartition, the degrees of freedom in the bulk will be equal to the surface degrees of freedom.
Now, we will see whether a dynamical dark energy model satisfies holographic equipartition.

\subsection{Quintessence models with $\omega>-1$}
Consider a dynamical dark energy model with an equation of state parameter which is always greater than $-1$. Since $\omega$ is greater than $-1$, even 
in the final stage, the bulk degrees of freedom will always less than the surface degrees of freedom as per equations (\ref{eqn:nsurf2}) and (\ref{eqn:nbulk3}).
Thus, these models will never achieve holographic equipartition.

\subsection{Quintessence models with $\omega\geq-1$}
Here we consider dark energy models with $\omega\geq-1$ and $\omega \to -1$, in the final stage. Since $\omega\geq-1$, the bulk degrees of freedom will never
exceed the surface degrees of freedom as per equations (\ref{eqn:nsurf2}) and (\ref{eqn:nbulk3}). Moreover, in the final stage as $\omega \to -1$, $N_{bulk}$
will be equal to $N_{surf}$, indicating the attainment of holographic equipartition. It is noteworthy that here the cosmic expansion could be interpreted
as a tendency for achieving holographic equipartition, since the universe attains the equality of degrees of freedom in the final stage.

\subsection{Phantom like models with $\omega<-1$}
Let us consider a model with equation of state parameter less than $-1$. Since, $\omega<-1$, the degrees of freedom residing in the bulk will exceed the surface degrees
of freedom as per equations (\ref{eqn:nsurf2}) and (\ref{eqn:nbulk3}). In this case the equality of the degrees of freedom will not have a special relevance
and the cosmic evolution couldn't be explained as a tendency for satisfying holographic equipartition.

From the above analysis, it is clear that the most stringent condition for the validity of Padmanabhan's law is the occurrence of the de Sitter epoch 
as the end phase. Even in varying dark energy models it is possible that, the dark energy density will soon achieve a constant value under 
asymptotic conditions. Quintessence models with $\omega \to -1$ in the final stage could satisfy holographic equipartition. But, it has to be mentioned that those models with $\omega<-1$, in any stage of evolution could not be motivated by the emergence
of space. Also, models with $\omega>-1$, in the final stage will not attain holographic equipartition.

\section{Thermodynamic motivation for the constraints on $\omega$}
In this section, we analyze the entropy evolution of the dynamical dark energy models. Our aim here is to find a thermodynamic motivation for the 
constraints imposed by the holographic equipartition, on the equation of state parameter.

Let us consider a dynamical dark energy model with an equation of state parameter $\omega$, as in the 
previous section. It was known that the entropy due to the other cosmic components of the universe is much less than the horizon entropy. 
Compared to horizon entropy, it lags about 35 orders of magnitude \cite{EL}.
So the total entropy of the universe will be approximately equal that of the horizon \cite{Pavon},
\begin{equation}
 S_T \sim S_H.
\end{equation}
According to the Gibbons-Hawking proposal \cite{GH}, the entropy of the cosmological horizon in units of $k_B$ is,
\begin{equation}
 S_H=\frac{A_H}{4 l_p^2}
\end{equation}
where $A_H=4\pi r_H^2$ is the area of the Hubble horizon with radius $r_H.$ Substituting the Hubble horizon radius, 
$r_H=1/H$ we get,
\begin{equation}
 S_H=\frac{\pi}{l_p^2 H^2}
\end{equation}
The rate of increase of entropy with respect to the cosmic time is then given by,
\begin{equation}
 \dot{S}_H=-\frac{2\pi}{l_p^2} \frac{\dot H}{H^3}.
\end{equation}
Using the Friedmann equation and the continuity equation, the above equation  could be written as,
\begin{equation}\label{sdot}
 \dot{S}_H=\frac{3\pi}{l_p^2 H} \left(1+\omega \right ).
\end{equation}
On taking the time derivative of ${\dot S}_H,$ we get,
\begin{equation}\label{sddot}
 \ddot{S}_H = \frac{3\pi}{l_p^2 H} \dot\omega -\frac{3\pi \dot H }{l_p^2 H^2} (1+\omega).
\end{equation}
According to the generalized second law the entropy of the universe should not decrease. For this, the rate of change entropy  must be positive. That is,
${\dot S}_H \geq 0$, for the validity of the generalized second law. Also, our universe will behave as an ordinary macroscopic system if it proceed to a 
state of maximum finite entropy in the long run which is possible only if, ${\ddot S}_H < 0$ as $t \to \infty.$

Now, we will consider the three kinds of dark energy models as in the previous case.
\subsection{Quintessence models with $\omega>-1$}
In these models, $\dot{S}_H$ will always be greater than $0$ as $\omega$ is always greater than $-1$ as per equation (\ref{sdot}). These models satisfy 
the generalized second law in the form ${\dot S}_H \geq 0$. But, it has to be mentioned that ${\dot S}_H$ will never tend to zero since $\omega$ is 
always greater than $-1$. Thus the entropy of such models will not attain a constant finite value even in the asymptotic limit. Hence there is no possibility
for such models to satisfy the maximum entropy principle.

\subsection{Quintessence models with $\omega\geq-1$}
Dark energy models with $\omega\geq-1$ always and $\omega \to -1$, when $t \to \infty$, satisfy the generalized second law in the form ${\dot S}_H \geq 0$.
Moreover, as $\omega$ tends to $-1$, ${\dot S}_H \to 0$, in the final stage indicating the attainment of equilibrium. Since $\omega \to -1$,
the second term of equation (\ref{sddot}) vanishes in the long run. Then, $\dot \omega$, being negative, these models satisfy the maximum entropy principle
and behave as ordinary macroscopic systems.

\subsection{Phantom like models with $\omega<-1$}
Since $\omega<-1$, ${\dot S}_H$ of these models will be negative during the evolution. These type of models violate even the generalized second law as
per equation (\ref{sdot}). Hence it is difficult to get a thermodynamic motivation for such models.

The above discussions support the constraints imposed by the holographic equipartition on the equation of state parameter from a thermodynamic point of 
view. Both the holographic equipartition and the maximum entropy principle demands an asymptotically de Sitter universe with $\omega\geq-1$. These facts
also points at the feasibility of describing a varying dark energy model through the law of emergence.

\section{Feasibility of describing a dynamical dark energy model using the law of emergence}
In this section, we explicitly prove the feasibility of describing a varying dark energy model through the law of emergence. Let us consider a cosmological model with dark energy and cold dark matter as the cosmic components. The dark energy component $\rho_{\Lambda}$ has an 
equation of state $\omega_{\Lambda}$ and the other component is assumed as the non relativistic matter $\rho_m$ with equation of state $\omega_m$.

 The Friedmann 
equation and the conservation law for a flat universe consists of these two components can be expressed as,
\begin{equation} \label{eqn:Fried1}
 3H^2 = 8\pi G \rho_m + 8\pi G \rho_\Lambda
\end{equation}
\begin{equation}\label{eqn:Fried2}
\dot{\rho}_m + 3 H \left(\rho_m+p_m + \rho_\Lambda+p_\Lambda \right)=-\dot{\rho}_{\Lambda}, 
\end{equation}
Now we will check the consistency of this cosmological model with the holographic equipartition law. As explained earlier the evolution of the universe 
can be explained as the tendency to satisfy the holographic equipartition, if it satisfy equation (\ref{eqn:dvdt1}). Here, the number of degrees of freedom on the 
Hubble sphere can be obtained from equation (\ref{eqn:nsurf1}). The bulk degrees of freedom is the sum of the degrees of freedom corresponds to matter 
 and the dark energy, 
 \begin{equation}
  N_{bulk}=N_{matter}+N_{de}
 \end{equation}
 Both $N_{matter}$ and $N_{de}$ can be calculated using the relation (\ref{eqn:NB2}). Accordingly $N_{matter}$ become, 
 \begin{equation}\label{eqn:nmatter2}
  N_{matter}= \frac{2 \left(1+3\omega_m\right) \rho_m V}{k_B T}
 \end{equation}
 and $N_{de}$ can be expressed as,
 \begin{equation}\label{eqn:nde2}
  N_{de}= \frac{2 \left(1+3\omega_{\Lambda} \right) \rho_{\Lambda} V}{k_B T}
 \end{equation}
 where we have taken $\epsilon=-1$ for matter and $\epsilon=+1$ for dark energy. Substituting equations (\ref{eqn:nsurf1}),(\ref{eqn:nmatter2}) and (\ref{eqn:nde2}) 
 into equation (\ref{eqn:dvdt1}), we get,
 \begin{equation}\label{eqn:dVdt2}
  \frac{dV}{dt}=l_p^2\left(\frac{4\pi}{l_p^2 H^2}+\frac{2\left(1+3\omega_m\right) \rho_m V}{k_B T} + \frac{2 \left(1+3\omega_{\Lambda} \right) \rho_{\Lambda} V}{k_B T} \right),
 \end{equation}
Here, $V=\frac{4\pi}{3H^3}$ and $ T=\frac{H\hbar}{2\pi k_B}$.
From the Friedmann equation, the matter density $\rho_m$ can be expressed as,
\begin{equation}\label{eqn:rhom1}
 \rho_m=\frac{3H^2}{8\pi G} - \rho_{\Lambda}
\end{equation}
Substituting equation (\ref{eqn:rhom1}) into equation (\ref{eqn:dvdt2}), we get
\begin{equation}\label{eqn:dvdt3}
 -\frac{4\pi \dot H}{H^4}=\frac{6 \pi}{H^2}+\frac{6\omega_m }{H^2}+\frac{16\pi^2 \rho_\Lambda (\omega_\Lambda-\omega_m)}{H^4}.
\end{equation}
This equation represents the equipartition law for the varying dark energy models.  Here we have wrote down the equipartition law by obtaining the 
l.h.s. and r.h.s. of the equation (\ref{eqn:dvdt1}) individually. The exact validity of the law in the present case can be readily checked by considering any one 
side of the above equation and by proving it to be identical with the other side. For instance, let us consider the l.h.s of the above equation, in which the 
prominent term is $\dot H,$ which can be written as,
\begin{equation}
 \dot H= \frac{4\pi G}{3H} \left(\dot \rho_m+\dot \rho_\Lambda \right).
\end{equation}
Then, with the help of continuity equation, the above equation 
can be recast in the form,
\begin{equation}
 \dot H = -4\pi G\left(\frac{3H^2(1+\omega_m)}{8 \pi G}+\rho_\Lambda(\omega_\Lambda-\omega_m)\right)
\end{equation}
which on multiplication by a factor $\frac{4 \pi}{H^4}$, 
become identical to the r.h.s of the equation(\ref{eqn:dvdt3}).
Thus the left hand side of the above equation can be reduced to the right hand side.
Now we can check whether the model satisfies the condition, 
\begin{equation}
 N_{surf}\geq \epsilon N_{bulk}.
\end{equation}
From equations (\ref{eqn:nsurf1}) and (\ref{eqn:NB2}), the above condition can be expressed as,
\begin{equation}
 \frac{4\pi}{l_p^2 H^2} \geq  \frac{-2\left(1+3\omega\right)\rho V}{k_BT},
\end{equation}
where $V$ is the Hubble volume. The term $\left(1+3\omega\right)\rho$ consists of contributions from both the non-relativistic matter and dark energy.
The above inequality impose a constraint on the equation of state,
\begin{equation}
 \omega \geq -1,
\end{equation}
where the equality corresponds to the end de Sitter phase at which the degrees of freedom are equal, i.e. $N_{surf}=N_{bulk}$.
This altogether shows that a varying dark energy
 model with a final de Sitter state is consistent with the holographic equipartition law. If $\omega<-1$, $N_{bulk}$ will exceed $N_{surf}$ and such a
 universe will not obey the holographic equipartition law. It is important to note that if this model satisfies the constraints imposed by the holographic
 equipartition; $\omega\geq-1$ always and $\omega \to -1$ in the long run, it will also behave as an ordinary macroscopic system that proceeds to a state 
 of maximum entropy.

\section{Conclusion}
The evolution of this universe could be described as the emergence of cosmic space with the progress of cosmic time. More specifically, the accelerated
expansion of the universe could be described as a tendency for satisfying holographic equipartition or as a tendency for equalizing the degrees of freedom
residing in the bulk to the degrees of freedom on the surface of the horizon. What is remarkable here is the fact that, this holographic equipartition
could not be achieved without a dark energy component. The universe with a pure cosmological constant will naturally evolve to a final state that satisfies
holographic equipartition. In an earlier work, we have explicitly proved the consistency of the standard $\Lambda CDM$ model of the universe with the 
holographic equipartition law. But, it is to be mentioned that, even if this simplest realization of dark energy, the cosmological constant is good in
explaining the recent acceleration of the universe, two of it's major drawbacks, the cosmological constant problem and the cosmic co-incidence problem,
motivates one to think about the varying dark energy models. Decaying vacuum models, Holographic dark energy models and phantom models are some among them
and many of them have succeeded in explaining the cosmic coincidence problem and the fine tuning problem. All these results motivates us to think about
the validity of Padmanabhan's holographic equipartition in the context of varying dark energy models.

In the present work, we have analyzed the consistency of the varying dark energy models with Padmanabhan's holographic equipartition. It is found that
the dark energy models that tend to a final de Sitter epoch with $\omega\geq-1$, satisfy holographic equipartition. It is also shown that the models with
$\omega$ always greater than $-1$, will never achieve holographic equipartition, while $N_{bulk}$ exceed $N_{surf}$ for those models
with $\omega<-1$. We have also presented a thermodynamic motivation for these constraints imposed by the holographic equipartition on $\omega$. It is
noteworthy that a dynamical dark energy model which obey the two constraints $\omega\geq-1$, always and $\omega \to-1$, in the long run satisfies the 
generalized second law and the maximum entropy principle, while the models with $\omega$ always greater than $-1$, will not evolve to a thermodynamic
equilibrium and those with $\omega<-1$ violate even the generalized second law. It has to be mentioned that both the holographic equipartition and the entropy
maximization demands an asymptotically de Sitter universe with an equation of state $\omega\geq-1$, rather than a pure cosmological constant. Also, we 
have explicitly proved the feasibility of describing a varying dark energy model through the law of emergence.

The phantom models with $\omega<-1$, will not satisfy holographic equipartition and these models will not lead to the maximization of entropy. But, it
should be noted that these phantom models that lead to the big rip also suffer from quantum instabilities \cite{CM,CJM}. 
It is worth mentioning that quintessence models  with $\omega$ always greater than $-1$, will not satisfy holographic equipartition and thus could not be motivated
by the law of emergence. It is important to note that those models will not attain a thermodynamic equilibrium. Our results point at the viability of the 
law of emergence in supporting the varying dark energy models

\end{document}